\documentclass[11pt]{article}  
\usepackage{menuproc}
\usepackage{cite}
\usepackage{epsfig}
\usepackage{amsmath,amssymb}
\begin{document}
%
%
%
\titlematter{Electromagnetic, weak, 
and strong interactions of light mesons}%
{P. Maris}%
{Dept. of Physics, North Carolina State University, 
 Raleigh,  NC 27695-8202, U.S.A.}%
{
The ladder-rainbow truncation of the set of Dyson--Schwinger equations
is used to study a variety of electroweak and strong processes
involving light mesons.  The parameters in the effective interaction
are constrained by the chiral condensate and $f_\pi$; the current
quark masses are fitted to $m_\pi$ and $m_K$.  The obtained
electromagnetic form factors are in good agreement with the data.
Also the weak $K_{l3}$ decay and the radiative and strong decays of
the vector mesons agree reasonably well with the data.  Finally, we
indicate how processes such as $\pi$-$\pi$ scattering can be 
described within this framework as well.}
%
%

\section{Introduction}
Our goal is to describe the hadrons and their interactions in terms of
their constituents, quarks and gluons, using the underlying theory,
QCD.  The set of Dyson--Schwinger equations [DSEs] form a useful tool
for this purpose~\cite{review}.  In rainbow-ladder truncation, they
have been successfully applied to calculate the masses and decay
constants of light pseudoscalar and vector
mesons~\cite{Maris:1997tm,Maris:1999nt}.  The dressed-quark
propagator, as obtained from its DSE, together with the meson
Bethe--Salpeter amplitude [BSA], form the necessary elements for
calculations of strong interactions in impulse approximation, such as
the $\rho \to \pi\pi$ decay.  For electroweak processes, such as the
electromagnetic form factors, radiative decays, and semileptonic
decays, one also needs the $q\bar q\gamma$ and $q\bar q W$ vertices.

\subsection{Dyson--Schwinger equations}
The DSE for the renormalized quark propagator in Euclidean space is
\begin{equation}
\label{gendse}
 S(p)^{-1} = i \, Z_2\, /\!\!\!p + Z_4\,m(\mu) + 
        Z_1 \int\!\!\frac{d^4q}{(2\pi)^4} \,
	g^2 D_{\mu\nu}(p-q) \, \textstyle{\frac{\lambda^i}{2}}
	\gamma_\mu \, S(q) \, \Gamma^i_\nu(q,p) \;,
\end{equation}
where $D_{\mu\nu}(k)$ is the dressed-gluon propagator and
$\Gamma^i_\nu(q;p)$ the dressed-quark-gluon vertex.  The most general
solution of Eq.~(\ref{gendse}) has the form 
\mbox{$S(p)^{-1} = i /\!\!\! p A(p^2) + B(p^2)$} and is renormalized 
at spacelike $\mu^2$ according to \mbox{$A(\mu^2)=1$} and
\mbox{$B(\mu^2)=m(\mu)$} with $m(\mu)$ the current quark mass.

Mesons are described by solutions of the homogeneous BSE 
\begin{equation}
 \Gamma_H(p_+,p_-;Q) = \int\!\!\frac{d^4q}{(2\pi)^4} \, 
        K(p,q;Q) \; S(q_+) \, \Gamma_H(q_+,q_-;Q) \, S(q_-)\, ,
\label{homBSE}
\end{equation}
at discrete values of $Q^2 = -m_H^2$, where $m_H$ is the meson mass.
In this equation, $p_+ = p + \eta Q$ and $p_- = p - (1-\eta) Q$ are
the outgoing and incoming quark momenta respectively, and similarly
for $q_\pm$.  The kernel $K$ is the renormalized, amputated $q\bar q$
scattering kernel that is irreducible with respect to a pair of $q\bar
q$ lines.  Together with the canonical normalization condition for
$q\bar q$ bound states, Eq.~(\ref{homBSE}) completely determines the
bound state BSA $\Gamma_H$.  Different types of mesons, such as
(pseudo-)scalar, (axial-)vector, and tensor mesons, are characterized
by different Dirac structures.

The dressed $q\bar q\gamma$ and $q\bar q W$ vertices satisfy an
inhomogeneous BSE: e.g. the quark-photon vertex
\mbox{$\Gamma_\mu(p_+,p_-;Q)$}, with $Q$ the photon momentum and
$p_\pm$ the quark momenta, satisfies~\cite{Maris:2000bh}
\begin{equation}
 \Gamma_\mu(p_+,p_-;Q) = Z_2 \, \gamma_\mu + 
        \int\!\!\frac{d^4q}{(2\pi)^4} \, K(p,q;Q) 
        \;S(q_+) \, \Gamma_\mu(q_+,q_-;Q) \, S(q_-)\, .
\label{verBSE}
\end{equation}
Solutions of the homogeneous version of Eq.~(\ref{verBSE}) define
vector meson bound states at timelike photon momenta
\mbox{$Q^2=-m_{\rm V}^2$}.  It follows that $\Gamma_\mu(p_+,p_-)$ has
poles at those locations~\cite{Maris:2000sk}.

\subsection{Model truncation}
\label{modelcalc}
To solve the BSE, we use a ladder truncation, 
\begin{equation}
        K(p,q;P) \to
        -{\cal G}\big((p-q)^2\big)\, D_{\mu\nu}^{\rm free}(p-q)
        \textstyle{\frac{\lambda^i}{2}}\gamma_\mu \otimes
        \textstyle{\frac{\lambda^i}{2}}\gamma_\nu \,,
\label{eq:ladder}
\end{equation}
in conjunction with the rainbow truncation for the quark DSE:
\mbox{$\Gamma^i_\nu(q,p) \rightarrow \gamma_\nu\lambda^i/2$} together
with \mbox{$Z_1 g^2 D_{\mu \nu}(k) \rightarrow {\cal G}(k^2)
D_{\mu\nu}^{\rm free}(k) $} in Eq.~(\ref{gendse}).  This truncation
preserves, independent of the details of the effective interaction
${\cal G}(k^2)$, both the vector Ward--Takahashi identity [WTI] for
the $q\bar q\gamma$ vertex and the axial-vector WTI.  The latter
ensures the existence of massless pseudoscalar mesons connected with
dynamical chiral symmetry breaking~\cite{Maris:1997tm}.  In
combination with impulse approximation, the former ensures
electromagnetic current conservation~\cite{Maris:2000sk}.

For the effective quark-antiquark interaction, we employ the Ansatz
given in Ref.~\cite{Maris:1999nt}.  The ultraviolet behavior of this
effective interaction is chosen to be that of the QCD running coupling
$\alpha(k^2)$; the ladder-rainbow truncation then generates the
correct perturbative QCD structure of the DSE-BSE system of equations.
In the infrared region, the interaction is sufficiently strong to
produce a realistic value for the chiral condensate of about
$(240\,{\rm GeV})^3$.  With this model, we can solve the BSE for
pseudoscalar and vector mesons, and calculate the meson masses and
leptonic decay constants.  The model parameters, along with the quark
masses, are fitted to give a good description of the chiral
condensate, $m_{\pi/K}$ and $f_{\pi}$.  The results of our model
calculations~\cite{Maris:1999nt} are shown in Table~\ref{tab:massdec}
and are in reasonable agreement with the data.  
\begin{table}[t]
\caption{Overview of results for the light pseudoscalar 
and vector meson masses and leptonic decay constants, all in GeV. 
Experimental data are from Ref.~\protect\cite{PDG}.
\label{tab:massdec}}
\begin{center}
\begin{tabular}{|l|cccc|cccccc|} \hline
			&$m_\pi$&$f_\pi$&$m_{K}$&$f_{K}$&
	$m_\rho$&$f_\rho$&$m_{K^*}$&$f_{K^*}$&$m_\phi$&$f_\phi$ \\
\hline
calc.   & 0.138 & 0.131 & 0.497 & 0.155 &
		0.742 & 0.207 & 0.936 & 0.241 & 1.072 & 0.259 \\ 
expt.	& 0.138 & 0.131 & 0.496 & 0.160 &
		0.770 & 0.216 & 0.892 & 0.225 & 1.020 & 0.236 \\
\hline
\end{tabular}
\end{center}
\end{table}
%

\section{Meson interactions}
In impulse approximation, processes such as electromagnetic
scattering, the weak $K_{l3}$ decay, radiative and strong decays of
vector mesons, can all be described by the same generic loop integral
\begin{equation}
  I^{abc}(P,Q,K) = N_c \int\!\!\frac{d^4q}{(2\pi)^4} \,
	{\rm Tr}\big[ S^a(q) \, \Gamma^{a\bar{b}}(q,q';P) 
        S^b(q') \, \Gamma^{b\bar{c}}(q',q'';Q) \, S^c(q'') \,
        \Gamma^{c\bar{a}}(q'',q;K) \big] \,, 
\label{eq:generictri}
\end{equation}
where $q - q' = P$, $q' - q'' = Q$, $q'' - q = K$, and momentum
conservation dictates $P + Q + K = 0$.  In Eq.~(\ref{eq:generictri}),
$S^i$ is the dressed quark propagator with flavor index $i$, and
$\Gamma^{i\bar{j}}(k,k';P)$ stands for a generic vertex function with
incoming quark flavor $j$ and momentum $k'$, and outgoing quark flavor
$i$ and momentum $k$.  Depending on the specific process under
consideration, this vertex function could be a meson BSA, a
$q\bar{q}\gamma$ vertex, or, in case of weak processes, a $q\bar{q}W$
vertex.  In the calculations discussed below, the propagators, the
meson BSAs, and the $q\bar{q}\gamma$ and $q\bar{q}W$ vertices are all
obtained as solutions of their respective DSE in rainbow-ladder
truncation, without adjusting any of the model parameters.

\subsection{Electromagnetic form factors}
Meson electromagnetic form factors in impulse approximation are
described by two diagrams, with the photon coupled to the quark and to
the antiquark respectively.  Each diagram corresponds to an integral
like Eq.~(\ref{eq:generictri}) with two meson BSAs and one
$q\bar{q}\gamma$-vertex.  With $Q$ being the photon momentum, and the
incoming and outgoing pseudoscalar mesons having momentum $P \mp Q/2$,
we can define a form factor for each of these
diagrams~\cite{Maris:2000sk}
\begin{eqnarray}
 	2\,P_\nu\,F_{a\bar{b}\bar{b}}(Q^2) &=&
			I_\nu^{abb}(P-Q/2,Q,-(P+Q/2)) \,.
\label{eq:emff}
\end{eqnarray}
We work in the isospin symmetry limit, and thus 
\mbox{$F_{\pi}(Q^2) = F_{u\bar{u}u}(Q^2)$}.  The $K^+$ and $K^0$
form factors are given by \mbox{$F_{K^+}= \frac{2}{3}F_{u\bar{s}u} + 
\frac{1}{3}F_{u\bar s \bar s}$} and \mbox{$F_{K^0} = 
-\frac{1}{3}F_{d\bar{s}d} + \frac{1}{3}F_{d\bar s\bar s}$} respectively.

\begin{figure}[t]
\parbox{.48\textwidth}{\centerline{\epsfig{file=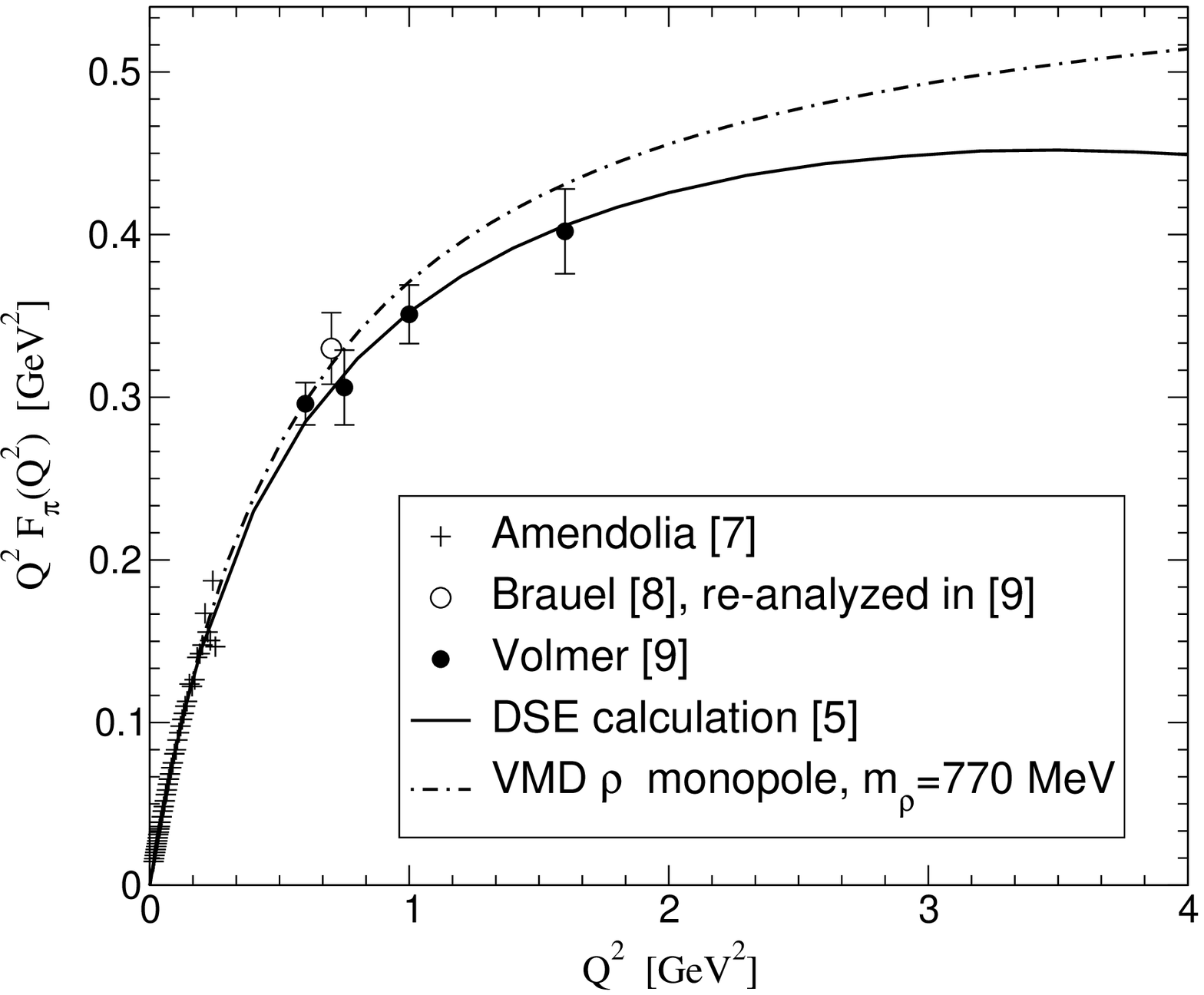,width=.45\textwidth,silent=,clip=}}}%
\hfill\parbox{.48\textwidth}{\centerline{\epsfig{file=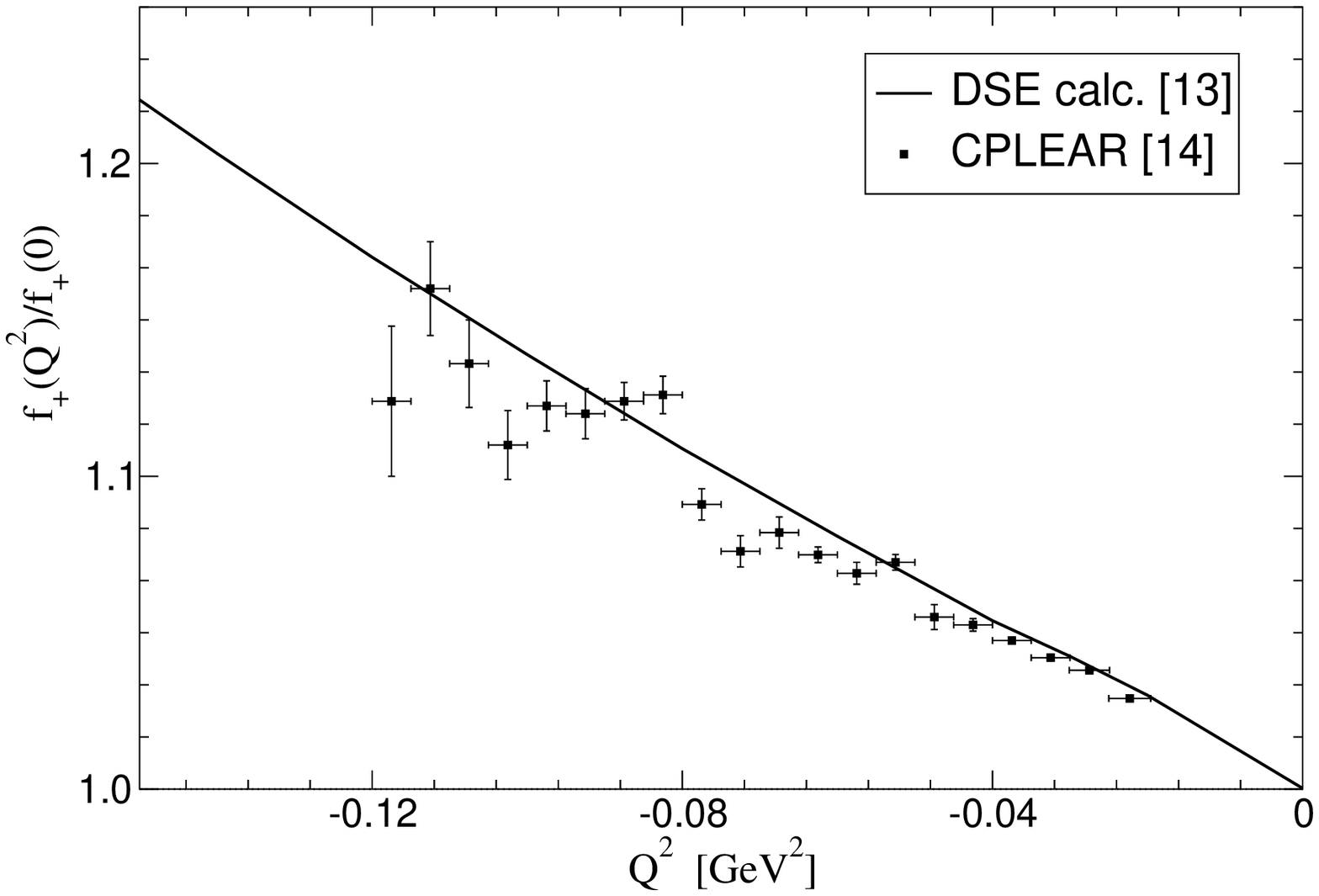,width=.45\textwidth,silent=,clip=}}}%
\caption{\label{fig:ewff}
On the left, our result for $Q^2 F_\pi(Q^2)$, and right, 
our curve for the $K_{l3}$ form factor $f_+(Q^2)$. }
\end{figure}
\begin{table}[b]
\caption{Calculated charge radii in ${\rm fm}^2$,
with expt. data~\protect\cite{A86,A86K,M78}, and $K_{l3}$ observables.  
The double entries for the expt. $K_{l3}$ data~\protect\cite{PDG} 
correspond to the neutral and charged $K_{l3}$ decays respectively.  
\label{tab:eweak}}
\begin{center}
\begin{tabular}{|l|ccc|ccccc|cc|} \hline
& $r^2_\pi$ 	& $r^2_{K^+}$ 	& $r^2_{K^0}$ 
& $\lambda_+$	& $\lambda_0$ 	& $-\xi$
& $\Gamma(K_{e3})$	& $\Gamma(K_{\mu3})$\\
\hline
calc.   & 0.45  & 0.38  & -0.086 
	& 0.027		& 0.018 	& 0.11 
	& 7.38$\cdot 10^6 s^{-1}$& 4.90$\cdot 10^6 s^{-1}$ \\
expt.	& 0.44  & 0.34  & -0.054 
	& .0276, .0288 	& .006, .025	& 0.31, 0.11 
	& 7.50, 3.89 & 5.26, 2.57 \\
\hline
\end{tabular}
\end{center}
\end{table}
Our result for $Q^2 F_\pi$ is shown in Fig.~\ref{fig:ewff}, together
with experimental data from Refs.~\cite{A86,Brauel:1979zk,Volmer01};
the corresponding charge radius, together with the neutral and charged
kaon charge radii, are given in Table~\ref{tab:eweak}.  The obtained
charge radii agree quite well with the experimental
data~\cite{A86,A86K,M78}, as do our form factors.  Up to about $Q^2 =
2\,{\rm GeV}^2$, our result for $F_\pi$ can be described very well by
a monopole with mass scale given by our calculated $m_\rho$,
$m=742\,{\rm MeV}$.  Above this value, our curve starts to deviate
more and more from this naive VMD monopole.  Our results for $F_K$ are
given in Ref.~\cite{Maris:2000sk} and can be fitted quite well up to
about $Q^2 = 2\sim3 \,{\rm GeV}^2$ by a monopole with mass scale
slightly larger than the $\rho$ mass.  Asymptotically, these form
factors behave like \mbox{$Q^2 F(Q^2) \rightarrow c$} up to
logarithmic corrections~\cite{Maris:1998hc}.  However, numerical
limitations prevent us from accurately determining the constants $c$.

\subsection{Weak interactions}
The matrix element $ \langle\pi^-(P+Q/2)|\bar{s}\gamma_\mu
u|K^0(P-Q/2)\rangle$ describing the semileptonic decay of neutral
kaons via a $W$-boson with momentum $Q$ can be characterized by two
form factors
\begin{eqnarray}
  I_{\mu}^{dsu}(P-Q/2,Q,-(P+Q/2))
	& = &  2 \, P_\mu \, f_+(Q^2) + Q_\mu \, f_-(Q^2)  \; .
\label{eq:fkl3}
\end{eqnarray}
The form factors $f_\pm$ for the $K^0$ decay are essentially the
same as those for $K^+$; in the isospin limit, the only difference
between the matrix elements for the $K^0$ and the $K^+$ decay is a
factor of $\sqrt{2}$, the $\pi^0$ being $(u\bar{u} -
d\bar{d})/\sqrt{2}$, which results in a factor of 2 difference in the
partial decay width.

In the right panel of Fig.~\ref{fig:ewff} we show our result for
$f_+(t)$~\cite{Ji:2001pj}, together with the experimental data for
this form factor~\cite{Apostolakis:2000gs}.  Experiments are often
characterized in terms of the transverse, $f_+$, and the scalar form
factor $f_0$, rather than $f_-$, which is defined by
\begin{eqnarray}
 f_0(Q^2) &=& \frac{Q_\mu \; 
	I_{\mu}^{dsu}\big(P-Q/2,Q,-(P+Q/2)\big)}{m_K^2 - m_\pi^2}
	= f_+(Q^2) - \frac{Q^2}{m_K^2 - m_\pi^2} f_-(Q^2)  \,.
\end{eqnarray}
The dimensionless slope parameter $\lambda$ for these form factors is
defined as $\lambda = -m_\pi^2 f'(0)/f(0)$; $\xi$ is defined as
$f_-(0)/f_+(0)$.  The partial decay width can be obtained by
integrating the decay rate, which depends on the lepton masses.  
Both the shape and the magnitude we obtain for these form factors
agree well with experiments, as can be seen from Table~\ref{tab:eweak}.

\subsection{Radiative decay of vector mesons}
We can describe the radiative decay of the vector mesons using the
same loop integral, Eq.~(\ref{eq:generictri}), this time with one
vector meson BSA, one pseudoscalar BSA, and one
$q\bar{q}\gamma$-vertex~\cite{Maris:2001rq} .  The on-shell value
gives us the coupling constant, which can be used to calculate the
partial decay width.  For virtual photons, we can define a form factor
$F_{VP\gamma}(Q^2)$, normalized to 1 at $Q^2 = 0$, which can be used
in estimating meson-exchange contributions to hadronic
processes~\cite{Tandy:1997qf,VODG95,Hecht:2001fr}.

In the isospin limit, the $\rho^0\,\pi^0\,\gamma$ and
$\rho^\pm\,\pi^\pm\,\gamma$ vertices are identical, and are given by
\begin{eqnarray}
\label{eq:rpgver}
 \frac{1}{3} \, I^{uuu}_{\mu \nu}(P,Q,-(P+Q)) & = &
  \frac{g_{\rho\pi\gamma}}{m_\rho} \; \epsilon_{\mu\nu\alpha\beta} 
	P_\alpha Q_\beta  \, F_{\rho\pi\gamma}(Q^2) \,,
\end{eqnarray}
where $P$ is the $\rho$ momentum.  The $\omega\,\pi\,\gamma$ vertex is
a factor of 3 larger, due to the difference in isospin factors.  For
the $K^\star \to K \gamma$ decay, we have to add two terms: one with
the photon coupled to the $\bar{s}$-quark and one with the photon
coupled to the $u$- or $d$-quark, corresponding to the charged or
neutral $K^\star$ decay respectively.

As Eq.~(\ref{eq:rpgver}) shows, it is $g_{VP\gamma}/m_V$ that is the
natural outcome of our calculations; therefore, it is this combination
that we report in Table~\ref{tab:vecdec}, together with the
corresponding decay widths~\cite{Maris:2001rq}.  The agreement between
theory and experiment for $g_{VP\gamma}/m_V$ is within about 10\%,
except for the discrepancy in the charged $K^\star \to K \gamma$ decay for which we have no explanation.  Likewise the large difference between
the neutral and charged $\rho$ decay width is beyond the reach
of the isospin symmetric impulse approximation.  Note that part of the
difference between the experimental and calculated decay width comes
from the phase space factor because our calculated vector meson masses
deviate up to 5\% from the physical masses.
\begin{table}[b]
\caption{
Vector meson radiative decays: coupling $g/m$ in GeV$^{-1}$ 
and partial decay width in keV.
\label{tab:vecdec}}
\begin{center}
\begin{tabular}{|l|ccc|cc|cc|cc|} \hline
& $g/m$ & $\Gamma_{\rho^\pm\pi^\pm\gamma}$ 	
& $\Gamma_{\rho^0\pi^0\gamma}$ 	
& $g/m$  & $\Gamma_{\omega\pi\gamma}$	
& $g/m$  & $\Gamma_{K^{\star\pm} K^\pm\gamma}$	
& $g/m$ & $\Gamma_{K^{\star 0} K^0\gamma}$ 	\\
\hline
calc.	& 0.69 & 53 &  (53) & 2.07 & 479 & 0.99	& 90 & 1.19 & 130 \\
expt.   & 0.74 & 68 & (102) & 2.31 & 717 & 0.83	& 50.3 & 1.28 & 116 \\
\hline
\end{tabular}
\end{center}
\end{table}

\subsection{Strong decays of vector mesons}
If we continue the calculation of the electromagnetic form factors
into the timelike region, we find a pole at the mass of the vector
meson bound states.  Using the behavior of the electromagnetic form
factors $F_{u\bar{u}u}$ and $F_{u\bar{s}\bar{s}}$ around this pole, we
can extract the coupling constants $g_{\rho\pi^+\pi^-}$ and $g_{\phi
K^+ K^-}$ respectively~\cite{Maris:2001rq} , which govern the strong
decays $\rho\rightarrow \pi\pi$ and $\phi\rightarrow K K$.  The
results from this analysis are given in Table~\ref{tab:strong}, and
are reasonably close to the experimental data.  Similarly, the two
form factors $f_+$ and $f_0$ describing the weak $K_{l3}$ decay
exhibit poles at $Q^2 = -m_{K^\star}^2$ and $Q^2 = -m_{\kappa}^2$
respectively, due to vector and scalar $u\bar{s}$ bound
states~\cite{Ji:2001pj}.  From the behavior close to the pole we can
extract the coupling constant for the strong decay $K^\star \to K\pi$
as well.  A direct calculation of the strong vector meson decays,
using on-shell meson BSAs but different numerical
techniques~\cite{JMT01vec}, agrees reasonably well with these results
extracted from the electroweak form factors.  Note that the factor of
three difference between the experimental and our calculated decay
width for the $\phi$ is due to the phase factor $(1 - 4
m_K^2/m_\phi^2)^{3/2}$: with our calculated masses, this factor is
$0.051$, whereas with the actual physical masses this factor is
$0.015$.  The dimensionless coupling constants agree within 10\% to
15\% with the experimental data.
\begin{table}[t]
\caption{
Overview of our results for vector meson strong decays (left):
dimensionless coupling constants and partial decay width in MeV,
and right, $\pi$-$\pi$ scattering lengths, compared to leading order 
chiral perturbation theory, $a_0^0 = 7 m_\pi^2/(8 \pi f_\pi^2)$ and
$a_0^2=-m_\pi^2/(4 \pi f_\pi^2)$~\cite{Weinberg:1966kf} (Weinberg's
limit).
\label{tab:strong}}
\parbox{0.62\textwidth}{
\begin{center}
\begin{tabular}{|l|cc|cc|cc|} \hline
& $g_{\rho\pi\pi}$	
& $\Gamma_{\rho\pi\pi}$	
& $g_{\phi KK}$	
& $\Gamma_{\phi KK}$	
& $g_{K^\star K\pi}$	 
& $\Gamma_{K^\star K\pi}$ \\	 
\hline
calc. & 5.4  & 115 & 4.3  &  6.7 & 4.0  & 31	\\
expt. & 6.02 & 151 & 4.64 &  2.2 & 4.60 & 50	\\
\hline
\end{tabular}
\end{center}}
\hfill\parbox{0.34\textwidth}{
\begin{center}
\begin{tabular}{|l|cc|} \hline
& $ a_0^0$	& $a^2_0$ \\
\hline
calc. 		& 0.170 & 0.045 \\  
Ref.~\cite{Weinberg:1966kf} & 0.156 & 0.045 \\
\hline
\end{tabular}
\end{center}}
\end{table}

\section{$\pi$-$\pi$ scattering}
Although impulse approximation seems to work remarkably well for a
variety of interactions involving three external particles, one has to
go beyond impulse approximation in order to describe processes with
four (or more) external particles.  As an example, consider
$\pi$-$\pi$ scattering at threshold.  The generic loop integral for
$\pi$-$\pi$ scattering in impulse approximation is
\begin{eqnarray}
 A &=& N_c \int\!\!\frac{d^4q}{(2\pi)^4} \,
{\rm Tr}\big[ S(q) \, \Gamma_\pi(q,q') \, S(q') \, 
	\Gamma_\pi(q',q'') \, S(q'') \,\Gamma_\pi(q'',q''') \, 
			S(q''') \,\Gamma_\pi(q''',q) \big] \,.
\label{eq:genericfour}
\end{eqnarray}
At threshold in the chiral limit, $\Gamma_\pi(k+P/2,k-P/2) \to i\gamma_5
B(k^2)/f_\pi$, and thus 
\begin{eqnarray}
  A &\to& 4 \, N_c \int\!\!\frac{d^4k}{(2\pi)^4} \,
        \frac{B^4(k)/f_\pi^4}{(k^2 A^2(k) + B^2(k))^2} \,,
\end{eqnarray}
which is nonzero.  On the other hand, chiral symmetry dictates that
the threshold scattering amplitudes vanish in the chiral limit like
$m_\pi^2/f_\pi^2$~\cite{Weinberg:1966kf}.  Clearly, impulse
approximation is insufficient to describe $\pi$-$\pi$ scattering.

\begin{figure}[t]
\centerline{\epsfig{file=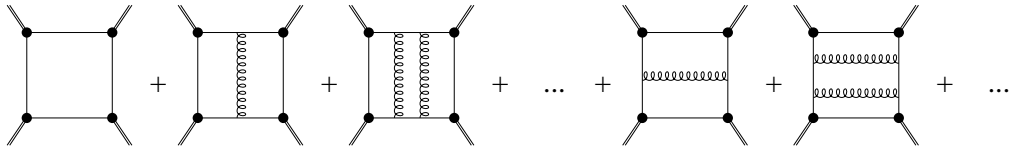,width=.85\textwidth,silent=,clip=}}%
\caption{\label{fig:pipiscat}
Diagrams needed to correctly describe $\pi$-$\pi$ scattering 
in rainbow-ladder truncation~\protect{\cite{Bicudo:2001jq}}.}
\end{figure}

In order to properly describe $\pi$-$\pi$ scattering in the
rainbow-ladder truncation, all possible diagrams with one or more
insertions of the ladder kernel $K$ across two pion BSAs should be
added to the impulse contribution~\cite{Bicudo:2001jq}, as indicated
in Fig.~\ref{fig:pipiscat}.
If we include these sets of ladder diagrams, we can show numerically
that the threshold $\pi$-$\pi$ scattering amplitudes indeed vanish
like $m_\pi^2/f_\pi^2$, using the same model as in the previous
section.  The corresponding scattering lengths are given in
Table~\ref{tab:strong}.  We expect that in particular $a_0^0$ will
receive significant corrections from pion loop effects, which we have
not included in our calculation: in chiral perturbation theory, higher
order corrections (i.e. pion loops) change the leading order result
to $a_0^0 = 0.220$ and $a_0^2 = 0.0444$~\cite{Colangelo:2000jc}.

So far, we have only considered $\pi$-$\pi$ scattering, but also 
in other hadronic 4-particle processes one should consider the
contributions from these infinite sums of ladder terms, in addition 
to the impulse term.  In general, we expect the role of these summed
ladder contributions to be less important than in $\pi$-$\pi$
scattering, except for processes that receive significant
contributions from resonances.  By adding these ladder diagrams one
can unambiguously incorporate $q\bar{q}$ bound state effects, and we
expect that this approach can provide a fundamental underpinning to
many processes described by effective meson lagrangians.


\acknowledgments{ I would like to thank Peter Tandy for our
collaboration on the electromagnetic form factors; the work on
$\pi$-$\pi$ scattering was done in collaboration with Pedro Bicudo,
Steve Cotanch, Felipe Llanes-Estrada, Emilio Ribiero, and Adam
Szczepaniak; I would also like to thank Dennis Jarecke, Chueng-Ryong
Ji, and Craig Roberts for useful discussions.  This work was funded by
DOE under grants No. DE-FG02-96ER40947 and DE-FG02-97ER41048, and by
NSF under grant No. INT-9807009, and benefitted from the resources of
the National Energy Research Scientific Computing Center.}


\end{document}